\documentclass{optica-article}

\journal{opticajournal} 

\articletype{Research Article}

\usepackage{lineno}
\usepackage{color}
\usepackage{xcolor}
\usepackage{CJK}

\modulolinenumbers[5]

\begin{document}
\begin{CJK*}{UTF8}{gbsn}
\title{Design and optimization of zone plates for flying focus applications}

\author{
Zhengkun Li,\authormark{1}
Yisong Zhou,\authormark{1}
Changbo Fu,\authormark{1,*}
Yugang Ma,\authormark{1}
}

\address{\authormark{1}Key Lab of Nuclear Physics and Ion-beam Application (MoE), Institute of Modern Physics, Fudan University, Shanghai 200433,  China\\
\email{\authormark{*}cbfu@fudan.edu.cn}
}



\begin{abstract*} 
Flying focus laser pulse technology, characterized by programmable velocity profiles and the ability to break the traditional link between focal spot motion and group velocity constraints, 
holds significant potential for revolutionary advances in high-intensity laser related fields.
However,  the flying focus generated with conventional approaches suffers from a low maximum intensity, as the energy is spread over a long distance, and the size of the focusing optics is also limited.
Here, we propose a zone-plate-based flying focus scheme in which the zone widths are modulated by a sinusoidal random distribution. 
Numerical simulations show that the high-order foci produced by the modulated zone plate can be suppressed by more than two orders of magnitude.
Meanwhile, the scheme sustains a high-intensity focal region extending over more than 100 Rayleigh lengths.
Furthermore, the focal speed can be tuned over a wide range, including values exceeding the speed of light in vacuum. 
Owing to these advantages, the proposed modulated zone-plate scheme offers new possibilities for high-intensity laser physics and holds promise for applications such as laser wakefield acceleration, photon acceleration, and more.


\end{abstract*}

\section{Introduction}

Spatiotemporal structured light, characterized by the nonseparability of its spatial and temporal field components\cite{Akturk_2010,Zhu:05,PhysRevA.99.023856}, is a key enabler for advanced optical technologies, including attosecond pulse sequences\cite{PhysRevLett.108.113904,wheeler_attosecond_2012}, terahertz wave generation\cite{toth_tilted_2023}, and the creation of flying-focus pulses\cite{Sainte-Marie2017,Froula2018,Pigeon2024}. 
The flying focus approach, as a special spatiotemporal light scheme, 
has been used to optimize many types of laser-based applications that require velocity matching or extended interaction lengths\cite{Trends2025}, including laser wakefield acceleration\cite{PhysRevLett.124.134802,Caizergues2020,PhysRevX.9.031044}, nonlinear Thomson scattering \cite{PhysRevE.105.065201}, and Compton scattering \cite{PhysRevA.103.012215,ComptonZhao2025}.
However, the fulfillment of this considerable potential requires the development of versatile and efficient optical configurations for flying focus generation.

There are many methods to realize flying focus. 
One conventional approach uses a chromatic diffractive lens to focus a frequency-chirped laser pulse, where different wavelengths are focused at different longitudinal positions and the chirp controls their arrival time\cite{Froula2018,Spatial_and_temporal_2019,PhysRevLett.120.225001}. Another method employs refractive lenses with longitudinal chromatism combined with a chirped pulse to achieve tunable flying focus\cite{Jolly:20}.  Controllable spatiotemporal structuring via spatial light modulators and 4f pulse shapers can also generate flying foci\cite{Kondakci:18,Kondakci2019,PhysRevA.99.023856}. 
Additionally, a technique combining an axiparabola\cite{oubrerie:hal-03630262,Smartsev:19} with a radial echelon enables ultrashort flying foci: the axiparabola focuses different radial zones to different axial positions in the far field, while the echelon adjusts their relative timing\cite{PhysRevLett.124.134802,Pigeon2024}.

However, these traditional methods have several limitations.
The chromatic-diffractive-lens-based approach is constrained by the physical dimensions of the focusing optics, which  restricts the achievable focal volume and intensity. For the method using refractive lenses with a chirped pulse, the effective focus range does not extend over one Rayleigh range\cite{Pigeon2024}.
The major theoretical limitation of the approach with the combination of spatial light modulators and 4f pulse shapers is that spatiotemporally confined Airy wave packets can only exist at the speed of light, making freely tunable flying focus velocities impossible\cite{Kondakci:18,Besieris:19}. 
For the design using an axiparabola with  an echelon, the echelon's structure (flat, concave, or convex) defines a specific flying focus velocity, making this design inherently difficult to adjust an otherwise freely tunable parameter\cite{Pigeon2024}. 

To overcome the limitation of  the reduced peak intensity, the idea of utilizing zone plates as chromatic diffractive lens is highly attractive. 
Various types of zone plates, which can be fabricated on quartz substrates with a high damage threshold\cite{GUO2018227,Said:95}, have been proposed to increase the peak intensities. The Fresnel zone plates (FZP), composed of alternating opaque and transparent circular rings, enhance the intensity. But the high-order foci are the primary cause of elevated background noise. 
Therefore,   suppression  of them is essential\cite{Choy:94}.
Another possible zone plate is the Gabor zone plate (GZP)\cite{Gabor:48, Rogers:50}.
GZP, whose transmittance along the radius is a continuous sinusoidal function, can suppress all high-order foci. But they are technologically impossible for conventional nanofabrication methods to fabricate.
Many researchers have attempted to approach GZP using various methods\cite{Tatchyn:85, PATT1990554, Beynon:92, Beynon:00, Greve:13}. 

However, many alternatives to GZP, such as the annulus-sector-element coded zone plate\cite{Wei2011,Zang2023} 
and Stagger arrangement zone plate\cite{Chen:13}, are incompatible with the flying focus regime owing to their lack of azimuthal symmetry.
These wavelength-specific designs fail when applied to the multi-wavelength composition of a chirped Gaussian pulse, leaving high-order foci unresolved. The method chosen to approach the GZP in the  chirped Gaussian pulse case is the modulated width zone plate (MWZP)\cite{ZHANG2018387}, a type of binary zone plate with azimuthal symmetry, to suppress all high-order foci.

In this study, to address the issue of low focus intensity,
we propose a zone-plate-based flying focus scheme, 
which  is designed to enhance focal intensity, 
suppress high-order foci, 
and achieve controllable propagation velocity throughout the focal region.
This novel approach introduces a sinusoidal modulation of the zone widths, structurally approximating the statistical transparency profile of  the GZP. 
Moreover, the focus performance improves significantly as the number of illuminated zones increases,  thereby enabling a well-defined single focus.

In the following sections,
we first introduce the concept of flying focus and its important characteristics for various laser-plasma interactions. 
Next, the modified Fresnel integrals and design of the MWZP are illustrated. Finally, numerical simulations of the zone-plate-based flying focus are presented, along with a discussion of the fulfillment of  the design objectives.

\section{Methods}

\subsection{Flying Focus scheme}

\begin{figure}[htbp]
\centering\includegraphics[width=1\linewidth]{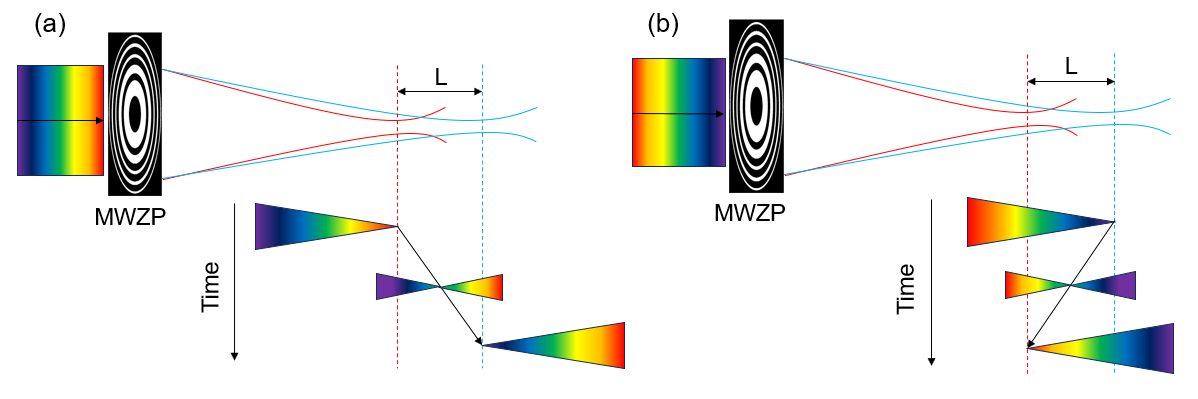}
\caption{The following are schematic diagrams illustrating: (a) a positively chirped laser beam and (b) a negatively chirped laser beam, both being focused by the MWZP. These setups result in the creation of an extended focal region. In the case of the positively chirped beam, the peak intensity travels (a) at a speed below the speed of light in the forward direction. Conversely, for the negatively chirped beam, the peak intensity can propagate (b) at a speed exceeding the speed of light in the backward direction\cite{Spatial_and_temporal_2019}.}
\label{flying focus concept}
\end{figure}

\begin{figure}[htbp]
\centering \includegraphics[width=0.75\linewidth]{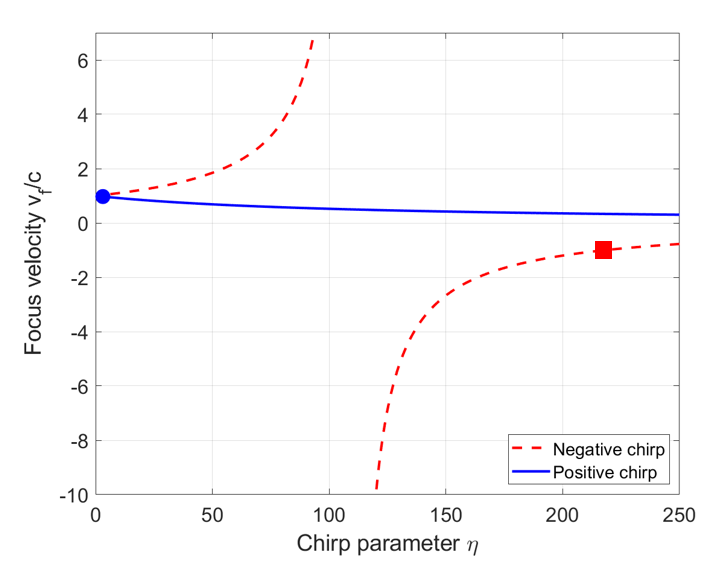}
\caption{The focal velocity of the flying focus pulse as a function of chirp parameter.
A positively chirped pulse is limited to positive subluminal focal velocities, while a negatively chirped pulse can have either a positive or negative velocity. The specific cases simulated in the following sections are marked with a blue circle (focal velocity $+0.97\mathrm{c}$) and a red square (focal velocity $-\mathrm{1.0c}$).
}
\label{chirp velocity}
\end{figure}

As illustrated in Figure \ref{flying focus concept}, the  flying focus scheme realized by a zone plate can be characterized by three fundamental properties that distinguish it from conventional focusing methods: a constant focal spot size, an extended focal range, and decoupling of focus velocity from the group velocity.
 First, the constant spot size implies that the instantaneous focal spot remains nearly diffraction-limited and maintains a consistently small diameter throughout the entire focal region.
 Second, the extended focal range indicates that the focal region spans a substantially longer longitudinal distance.
Most importantly, the decoupling of focus and group velocity allows the focal velocity to become a freely tunable parameter, enabling propagation at subluminal, luminal, or superluminal speeds.

The flying-focus constant spot size  sustains relatively high intensities over the focus range.
The analytical expression for the spatial evolution of the flying-focus beam radius along the propagation direction 
z is given for each spectral component of the pulse\cite{Spatial_and_temporal_2019}:
\begin{equation}
w _ { \lambda } ( z ) = R \Biggl [ \left( \frac { ( z + f _ { 0 } ) \lambda } { \pi R ^ { 2 } } \right) ^ { 2 } + \left( 1 - \frac { ( z + f _ { 0 } ) \lambda } { \lambda _ { 0 } f _ { 0 } } \right) ^ { 2 } \Biggr ] ^ { 1 / 2 },
\end{equation} where $\mathrm{R}$ represents the radius of the incident laser beam on the zone plate, $\mathrm{f}_0$ denotes the focal length corresponding to the central wavelength $\lambda_0$ in vacuum, and $\lambda(t)$ characterizes the time-dependent variation of the laser wavelength. The minimum spot size for light in the spectral range is $\mathrm{w}_0 \approx \frac{\lambda_0 \mathrm{f}_0}{\pi \mathrm{R}}$. In the extended focus range, the focal spot maintains a self-similar profile.

The long focal length is determined by the chromatic dispersion of
the optical system and spectral bandwidth of the input laser pulse. For the dispersion characteristics of the zone plates, red light with a longer wavelength will focus closer to the zone plate, whereas blue light with a shorter wavelength
will focus further away from the zone plate.
The range of foci, which is determined by the laser bandwidth ($\Delta\lambda = \lambda_{\mathrm{r}} - \lambda_{\mathrm{b}}$), provides the longitudinal focal shift $\Delta \mathrm{z} \simeq \mathrm{f}_{0} \Delta\lambda / \lambda_{0}$.

Maintaining a high intensity over an extended propagation range is fundamental to the success of key laser-plasma applications.  
For example, in dephasingless laser-wakefield acceleration\cite{Caizergues2020,PhysRevLett.124.134802,Miller2023}, reaching TeV energies in a 4.5-meter stage requires sustaining a nonlinear intensity (e.g., $>2.14 \times 10^{18}$  $\mathrm{W/cm}^2$ for 800 nm laser) throughout the entire distance. This requirement is equally vital in a photon accelerator\cite{PhysRevLett.61.337,PhysRevLett.123.124801}, where the final frequency upshift is directly related to the distance photons reside within an ionization gradient, necessitating that a strong peak intensity be maintained over long ranges for efficient conversion to extreme-ultraviolet light.

One of the key parameters characterizing the flying focus is its focal velocity.
The focal velocity $\mathrm{v}_\mathrm{f}$ denotes the rate at which the point of peak intensity propagates, a continuous progression driven by the chromatic focusing and the temporal separation of frequencies within the pulse\cite{PhysRevA.97.033835}:
\begin{equation}
\frac{v_f}{c} = \left( \frac{2f}{\eta c \tau^2 \omega_0} \right) \left( 1 + \frac{2f}{\eta c \tau^2 \omega_0} \right)^{-1},
\end{equation}
where $\tau$ is the transform-limited incident pulse duration, $\eta$ is the chirp parameter, and f is the focus of the zone plate. As illustrated in Figure \ref{chirp velocity},  the propagation velocity of the flying focus is highly tunable via the chirp parameter.   When dealing with a positively chirped laser pulse, the peak intensity propagates at a subluminal speed in the forward direction. The focus of the longest wavelength is initially formed at the leading edge of the focal region, as shown in Figure \ref{flying focus concept}a, while the light with the shortest wavelength reaches its focal spot last in time. By contrast, when a negatively chirped laser pulse is introduced
into the system, the resulting laser focus either co-propagates or counter-propagates with an arbitrary velocity. For the
counter-propagation case, the peak intensity can proceed with superluminal speeds easily. As shown in Figure \ref{flying focus concept}b, the light with the shortest wavelength focuses first at the focal point farthest from the zone plate. Subsequently, light with longer wavelengths comes into focus progressively closer to the zone plate. In further specific simulations, the chirp parameter is chosen to be 3.1 and -220, marked by the blue circle and red square, respectively, in Figure \ref{chirp velocity}. These chirp values yield focus velocities close to the speed of light in the forward or backward directions.

Appropriate focal velocities are important for experimental applications. In dephasingless laser-wakefield acceleration, applying an appropriate positive chirp to propagate the intensity peak at the vacuum speed of light is essential. This prevents the electrons from outrunning the accelerating phase. In a photon accelerator, employing the right negative chirp to generate an intensity peak that counterpropagates at vacuum light speed is fundamental. This avoids the photons from outpacing the ionization front.

\subsection{Design of zone plates for flying focus}

 To design and optimize zone plates for flying focus purpose,
the modified Fresnel integrals approach\cite{PhysRevA.97.033835}
will be employed.
The chirped Gaussian pulse, as well as the plane wave will be used as the input.
Three types of zone plates, MWZP, GZP, and FZP, will be included in the propagation process.
The intensity distributions on the output planes are simulated based on the  modified Fresnel integrals and the conventional angular spectrum method.

\subsubsection{The modified Fresnel integrals for flying focus}
The modified Fresnel integrals can be considered as a convolution of the Gaussian light beams passing through the zone plate and the modified impulse response function. 
In the convolution theory, the inverse Fourier transform of the product of two functions' Fourier transforms yields their convolution. 

To derive the transverse components' equation of the chirped Gaussian pulse, 
we simplify the incident light pulse as a plane wave with a modulating envelope:
$\mathrm{E}=\frac{1}{2}\mathrm{E}_0(\mathrm{r},\xi,\mathrm{z})\mathrm{exp}(\mathrm{i}(\mathrm{k}_0\mathrm{z}-\omega_0 \mathrm{t}))+c.c.$, 
where $\xi=\mathrm{ct-z}$ is the moving frame coordinate,  $\omega_0$ is the central frequency,
the wave number $\mathrm{k}_0=\omega_0 /\mathrm{c}$, 
and c is vacuum light speed. 
The $\xi$ domain integral is performed to obtain the transverse components of the envelope $\mathrm{E}_{\perp}$\cite{PhysRevA.97.033835},

\begin{equation}
    E_{\perp}(r,\xi,z)=-\frac{ik_{0}-\partial_{\xi}}{2\pi z}\int \exp\left[\frac{ik_{0}}{2z}(r-r')^2\right] \cdot E_{\perp}\left(r',\xi-\frac{|r-r'|^2}{2z},0\right)dr',
     \label{transverse electric fields}
\end{equation}
where $\mathrm{r}$ signifies the location for calculating the diffraction intensity, $\mathrm{r}'$ represents the incident zone plate whose intensity is already known, and z is the propagation distance on the optical axis. $\mathrm{E}_{\perp}(\mathrm{r}',\xi,0)$ determine the electric field's amplitudes and phases at $\mathrm{z}=0$; and the term $\frac{|\mathrm{r}-\mathrm{r}'|^2}{2\mathrm{z}}$ translating $\xi$ term considers the time differences of light propagating from different points on the designed zone plate. 
The requirement of the paraxial approximation implied by the previous equation is that the f-number should be greater than four. The f-number of the designed zone plate used in our investigation can easily satisfy the paraxial approximation.

 The transverse  electric fields of  the chirped Gaussian pulse are then split as a convolution of the Gaussian light beams with impulse response functions. The following assistant parameter, which contains the chirp term, is defined as,
\begin{equation}
A = c^2 \tau^2 (1 - j \cdot \eta).
\end{equation}
Here is the complete transfer function in the modified Fresnel integrals,
\begin{equation}
h\left(x_1,y_1\right) = \frac{1}{2 \pi z} \left( -j k_0 - \frac{2 \xi}{A} + \frac{x_1^2 + y_1^2}{Az } \right) \cdot \exp\left( \frac{j k_0}{2 z} (x_1^2 + y_1^2) + \frac{\xi (x_1^2 + y_1^2)}{A z} - \frac{(x_1^2 + y_1^2)^2}{4 A z^2} \right).
\end{equation}
The beam part of the convolution is then determined to be,
\begin{equation}
U_1 = t_{plate} \cdot E_0 \cdot \exp\left(-\frac{x_1^2 + y_1^2}{w_0^2} - \frac{\xi^2}{A}\right),
\label{input_beam}
\end{equation}
which is a Gaussian wave with $\xi$ term that passes through the zone plate.
\subsubsection{Design of the MWZP}
\begin{figure}
    \centering
    \includegraphics[width=1\linewidth]{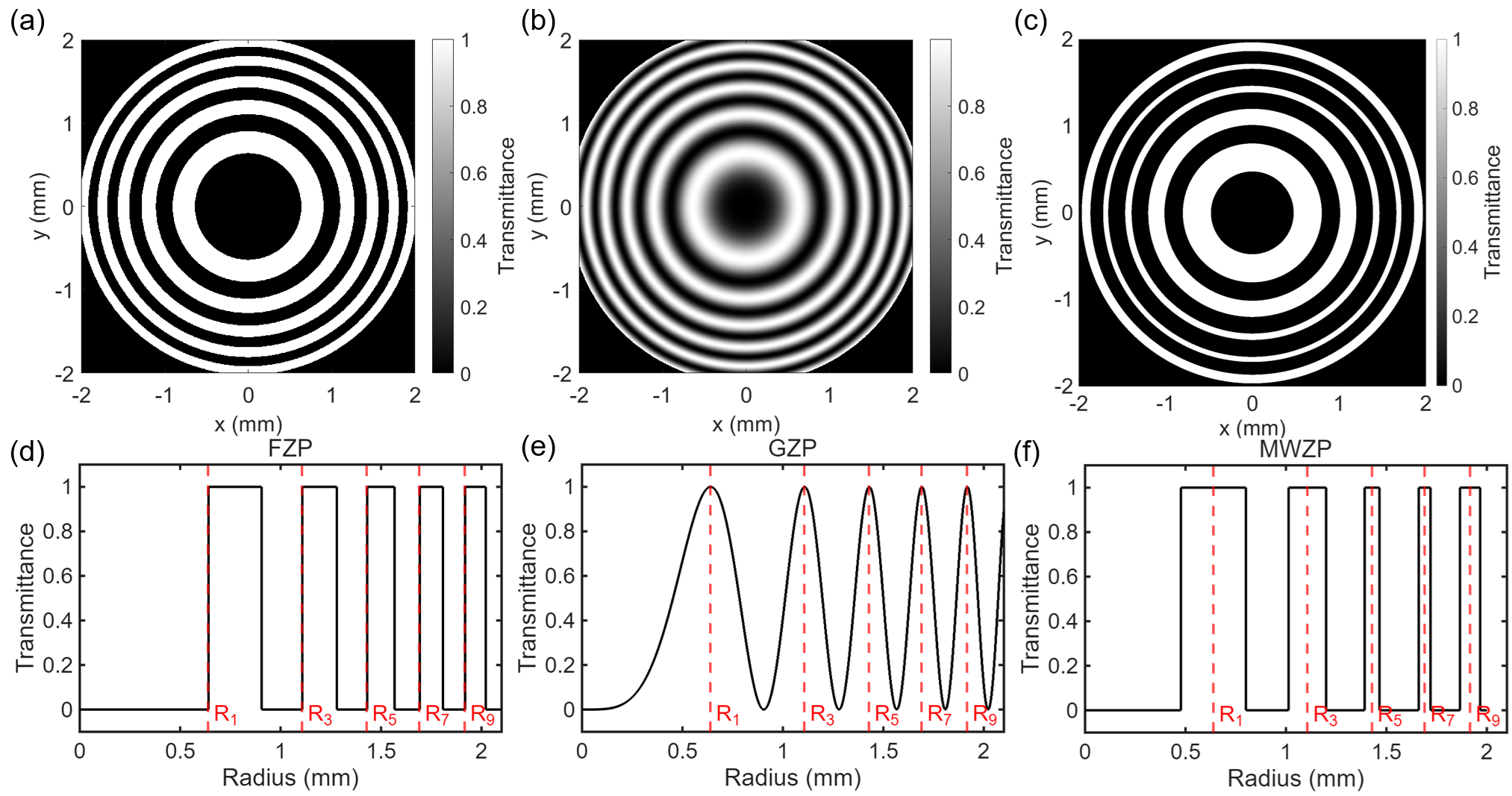}
    \caption{The comparison between FZP, GZP, and MWZP. (a) An FZP with 10 zones. (b) A GZP with 10 zones. (c) An MWZP with 10 zones. (d) The graph depicting the transmittance curve of the FZP as a function of its radius. (e) The graph showing the transmittance curve of the GZP versus its radius. (f) The graph representing the transmittance curve of the MWZP in relation to its radius\cite{ZHANG2018387}.}
    \label{zone plate design}
\end{figure}

The MWZP is designed by combining an FZP and a GZP.
Typically, an FZP provides a relatively high peak intensity at its primary focus but also produces multiple higher-order foci.
In contrast, a GZP does not generate higher-order foci; but it is more challenging to fabricate because of its continuous sinusoidal transmittance. 
The combined structure,  MWZP, is expected not only to suppress all high-order foci, but also to simplify the nanofabrication process by functioning as a binary zone plate.

The FZPs are composed of alternating opaque and transparent half-wave plates, as shown in Figure \ref{zone plate design}a. It is known that the radius of the nth zone is \cite{Hecht1987OpticsB},
\begin{equation}
    R_n = \sqrt{n f \lambda},
\end{equation} where $\lambda$ is the wavelength of the incident light, and f is the first-order focal length of the zone plate. The appropriate condition for this equation is 
 $ \mathrm{f} \gg \mathrm{n} \lambda $, which is satisfied in this study. The zones from the FZPs have equal areas $ \pi \lambda \mathrm{f} $. By using vibration vectors to accumulate contributions from all half-wave plates, the light fields can be significantly increased, which can be seen as a diffractive lens. The focus f is inversely proportional to the incident wavelength\cite{Hecht1987OpticsB},
 \begin{equation}
    f = \frac{\rho_1^2}{\lambda},
\end{equation} where $\rho_1$ is the radius of the first-zone plate. The transmission function of the FZP is a binary function plotted against radius, as shown in Figure \ref{zone plate design}d. The red dashed lines mark the radii of the various zones. The transmittance can be expressed in the following mathematical form \cite{ZHANG2018387},
\begin{equation}
t_F(r) = 
\begin{cases} 
0, & 2n\lambda f \leq r^2 < (2n + 1)\lambda f, \\
1, & (2n + 1)\lambda f \leq r^2 < (2n + 2)\lambda f, 
\end{cases}
\end{equation} where n is an integer. Here, the odd half-wave plates are set as transparent.

The GZPs' transmission function is a continuous sinusoidal function, which is distinct from FZPs' stair-like zones, as illustrated in Figure \ref{zone plate design}b. GZP's transmission function is expressed as follows\cite{Beynon:92},
 \begin{equation}
t_G(r) = \frac{1}{2} \left( 1 \pm \cos 2\pi \frac{r^2}{\alpha \rho_1^2} \right), 
\end{equation} where the focal length is $\mathrm{z}=\pm \frac{\alpha \rho_1^2}{2 \lambda}$. In general, $\alpha$ is a positive integer, chosen to be two. Consequently, the focal length of the GZP is determined to be $\frac{\rho_1^2}{ \lambda}$, which is the same as that of FZP. The sign in the Gabor transmission equation is determined to be negative, which defines a negative GZP. Its continuous transmittance between 0 and 1 is also displayed in Figure \ref{zone plate design}e. The red dashed lines indicate the central maxima of  different zones. Owing to the continuous transmission line, there exists a technological impossibility of physically approaching the GZP with current nanofabrication technology. Over the last three decades, various approaches have been used to imitate the sinusoidal transmission function of the GZPs.

Following one of the imitation methods of GZPs\cite{ZHANG2018387}, an MWZP is constructed by changing the width of every Fresnel half-wave plate, as shown in Figure \ref{zone plate design}c. Firstly, to avoid fabrication impossibility, the continuous zones of the standard GZPs are replaced with zone pairs of a binary transmission function, with the center of zone pairs $\mathrm{R}_{2\mathrm{n}-1}$ unchanged. Secondly, the width of the transparent ring is modulated with a sinusoidal distribution: before replacement, the zone pair width is $\mathrm{R}_{2\mathrm{n}}-\mathrm{R}_{2\mathrm{n}-2}$; after replacement, the nth transparent zone width of the MWZP is\cite{ZHANG2018387},
\begin{equation}
d_n=2(R_{2n}-R_{2n-1})\cdot s_n,
\end{equation} where $\mathrm{s}_\mathrm{n}$ denotes the series of random variables. A single s in series $\mathrm{s}_\mathrm{n}$ satisfies the following probability density function\cite{ZHANG2018387}, 
 \begin{equation}
\rho(s) = 0.5\pi \cos\left[\pi (s - 0.5)\right] \cdot Rect(s-0.5).
\label{density_function}
\end{equation} 
Here ``Rect'' is a rectangle function, 
\begin{equation}
\mathrm{Rect}(s - 0.5) = 
\begin{cases} 
1, & 0 \leq s < 1 \\
0, & \text{otherwise.}
\end{cases}
\end{equation}
The  \(\rho(\mathrm{s})\) is normalized, i.e. \(\int_0^1 \rho(\mathrm{s}) \mathrm{ds} = 1\). 
The cumulative distribution function F(s) is then obtained by integrating the probability density function $\rho(\mathrm{s})$\cite{Feller1971Probability}, 
\begin{equation}
F(s) = \int_{0}^{s}\rho(s)ds=0.5 \sin \left[ \pi (s - 0.5) \right] + 0.5.
\end{equation} 
And then, $\mathrm{s}$ can be calculated inversely as,
 \begin{equation}\label{eq.sin.random}
s = F^{-1}(s)= 0.5 + \frac{1}{\pi} \arcsin(2u - 1),
\end{equation}
where u is between 0 and 1. 
Modulating the zone width in this manner results in an MWZP (Figure \ref{zone plate design}c), whose transmittance function is plotted against the radius in Figure \ref{zone plate design}f. 
The MWZP can be described as a step function\cite{ZHANG2018387}, 
 \begin{equation}
t_{MWZP}(r) = 
\begin{cases} 
1, & R_{2n-1} - \frac{d_n}{2} \leq r < R_{2n-1} + \frac{d_n}{2} \\
0, & \text{otherwise},
\end{cases}
\end{equation} 
where n is an integer. 
 As a result of the width modulation in Eq.\ref{density_function}, the expectation value of the transmission probability for the incident light at a certain radius is approximately the transmission probability of GZPs, which results in similar diffraction features. Compared with other binary GZPs, there are no microstructures in this MWZP design\cite{Beynon:92,Beynon:00,Greve:13}, which is beneficial for future fabrication with an increased number of zones.

\subsubsection{The design's objective for flying focus scheme}

Applying the modified Fresnel integrals to the MWZP, this study aims to generate the flying focus with  \emph{no high-order foci with reduced background noise}, \emph{enhanced peak intensities}, and \emph{controllable focus velocities across the extended focal range}.
A schematic of the setup is presented in Figure \ref{flying focus concept}, 
which illustrates that a chirped Gaussian pulse passes through an MWZP to generate a lengthened focal region.

\section{Results}

Unless otherwise specified, the following parameters are used in the simulation.
The incoming laser has a central wavelength of 800 nm,  pulse duration of 115 fs, peak power of 361.6 MW, and chirp parameters of 3.1 and -220.
For the zone plates simulated here, 
three types of zone plates, MWZP,GZP, and FZP
are considered, which have  primary focal lengths of 51 cm.
The input planes, output planes, and zone plates are all discretized into $7800\times 7800$ pixels.
The number 7800 is chosen to qualify the stringent standard $\frac{\mathrm{k}_0\mathrm{w}_0}{\pi \mathrm{f}^{\#}}$\cite{PhysRevA.97.033835}, where $\mathrm{f}^{\#}$ is the f-number of the zone plate considered.
Table \ref{parmeter table1}  lists the major parameters used in the numerical simulations.

The conventional angular spectrum method\cite{Goodman1969IntroductionTF}
is employed to simulate the  plane wave propagation through the zone plates;
while for { the chirped Gaussian pulse}, the modified Fresnel integrals  described in Section 2.2.1\cite{PhysRevA.97.033835} is used.
The output beam intensity  is described by the function $\mathrm{I}_{\mathrm{N}}^{\mathrm{ZT}}(\mathrm{x},\mathrm{y},\mathrm{z},\mathrm{t})$, where the superscript $\mathrm{ZT}$ indexes the zone-plate type,
and the underscript $\mathrm{N}$ signifies the total number of zones used.
By modulating zone plate width  $\mathrm{d}_\mathrm{n}$, 
zone number N, 
and the chirp parameter  $\eta$, 
the corresponding output intensity $\mathrm{I}_{\mathrm{N}}^{\mathrm{ZT}}(\mathrm{x},\mathrm{y},\mathrm{z},\mathrm{t})$ can be obtained.

\begin{table}[htbp]
\caption{Parameters used in numerical simulations.}
  \label{tab:shape-functions}
  \centering
\begin{tabular}{cccc}
\hline
Type& Parameters & Value \\
\hline
 Chirped  Gaussian Pulse parameters & wavelength $\lambda (nm)$ & 800  \\
& $\Delta\lambda (nm)$ & 19  \\
&$\tau(fs)$ & 115 \\
& $P_{peak}(MW)$ & 361.6 \\
&chirp parameter $\eta$& 3.1 and -220\\
Zone plate parameters &The first zone radius $\rho_1 (\mu m)$ & $638.75$  \\
&Focal length $f_1(cm)$ & $51$  \\
&Maximum considered $w_0$(cm) & $2.8$  \\
&Pixels in a transverse dimension  & $7800$  \\
\hline
\end{tabular}
\label{parmeter table1}
\end{table}

\subsection{Modulating zone plate width}

\begin{figure}[htbp]
\centering \includegraphics[width=1\linewidth]{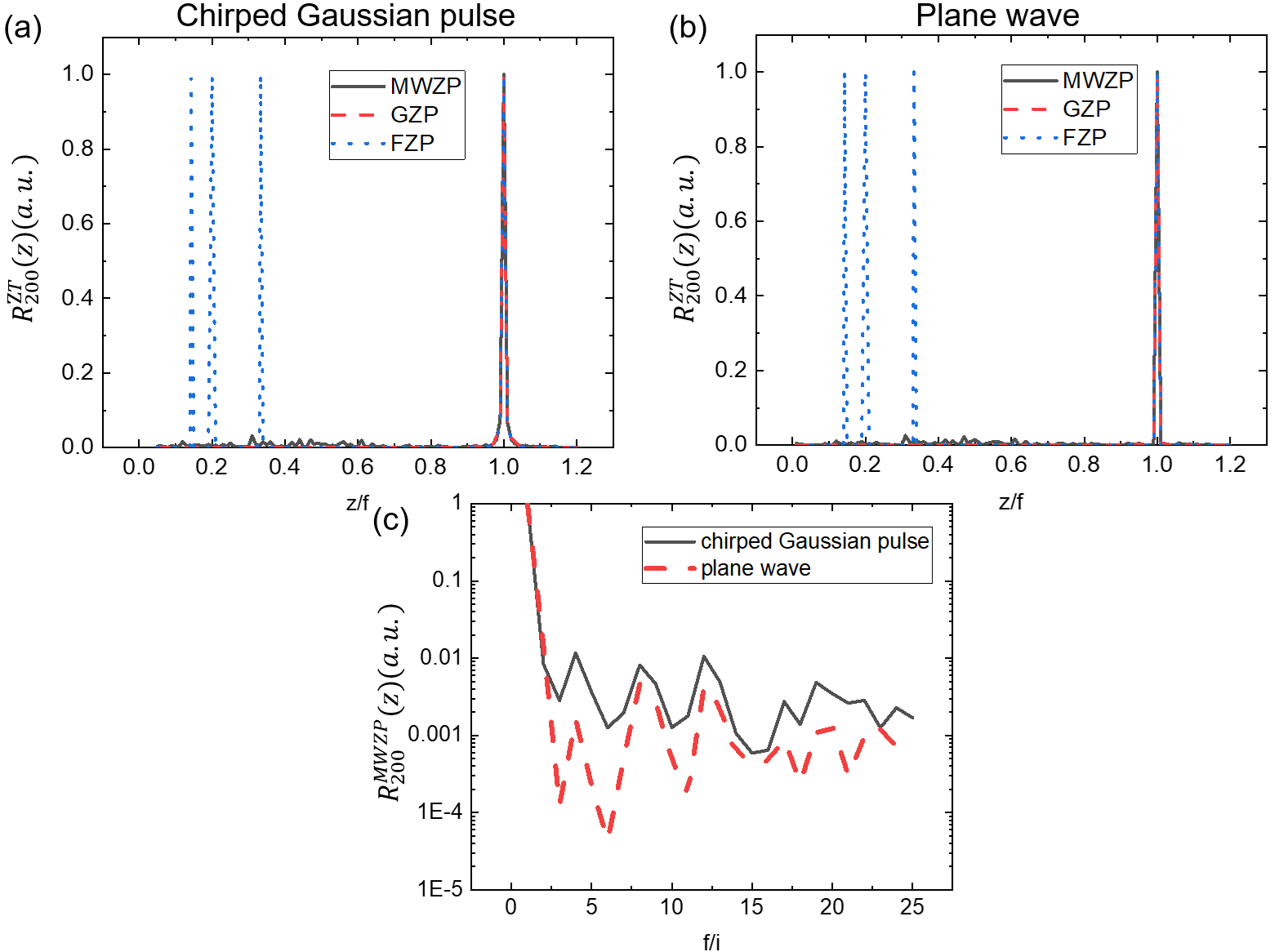}
\caption{Simulated on-axis normalized diffraction intensity distribution $\mathrm{R}_{200}^{\mathrm{ZT}}(\mathrm{z})$ (in units of z/f) for (a) a chirped Gaussian pulse and (b) monochromatic plane wave propagation. The profiles compare the performance of an MWZP with 200 zones (black solid line) against a GZP (200 zones, red dashed line) and an FZP (200 zones, blue dotted line). (c) The normalized intensity of the MWZP $\mathrm{R}_{200}^{\mathrm{MWZP}}(\mathrm{z})$ at discrete high-order focal positions f/i (where i is an integer) is distinctly shown for both propagation cases. 
}
\label{comparison_3plates_revision}
\end{figure}

Taking $u$ in Eq.\ref{eq.sin.random} as a random number between 0 and 1,  an MWZP is analyzed, together with a GZP and an FZP.
The relative intensity, $\mathrm{R}_{\mathrm{N}}^{\mathrm{ZT}}(z)$, can be defined as,
\begin{equation}\label{eq.Rn.zt}
    R_{N}^{ZT}(z)=\frac{I_{N}^{ZT}(x=0,y=0,z,t=\frac{z}{c})}{I_{N}^{ZT}(x=0,y=0,z=f_1,t=\frac{f_1}{c})},
\end{equation}
where $\mathrm{f}_1$ is the primary focal length.

Figure \ref{comparison_3plates_revision}a compares the on-axis normalized  intensity $\mathrm{R}_{200}^{\mathrm{ZT}}(\mathrm{z}) $ of the MWZP,  GZP, and  FZP, 
 where the input beam is a chirped Gaussian pulse.
One can find that the ratios of the intensities of  the high-order foci  and the main focus  intensities of the GZP and MWZP are suppressed.
However, for the FZP, the 3rd, 5th, and 7th-order foci still exist, which have almost equal strength as the primary focus. 

For comparison, 
if the input beam  is a monochromatic plane wave, 
instead of a chiped Gaussian pulse,
the results are shown in Figure \ref{comparison_3plates_revision}b. The results are almost the same as those in the chirped Gaussian pulse case.

Figure \ref{comparison_3plates_revision}c illustrates the $\mathrm{R}_{200}^{\mathrm{MWZP}}(\mathrm{z})$, where the input beams are a plane wave (dashed line) and a chirped Gaussian pulse (solid line). 
Even for a chirped Gaussian pulse which involves multiple wavelengths, 
the relative intensities at  high-order focal spots can still be depressed more than 100 times. 

\subsection{Modulating zone plate number}

\begin{figure}[htbp]
\centering \includegraphics[width=1\linewidth]{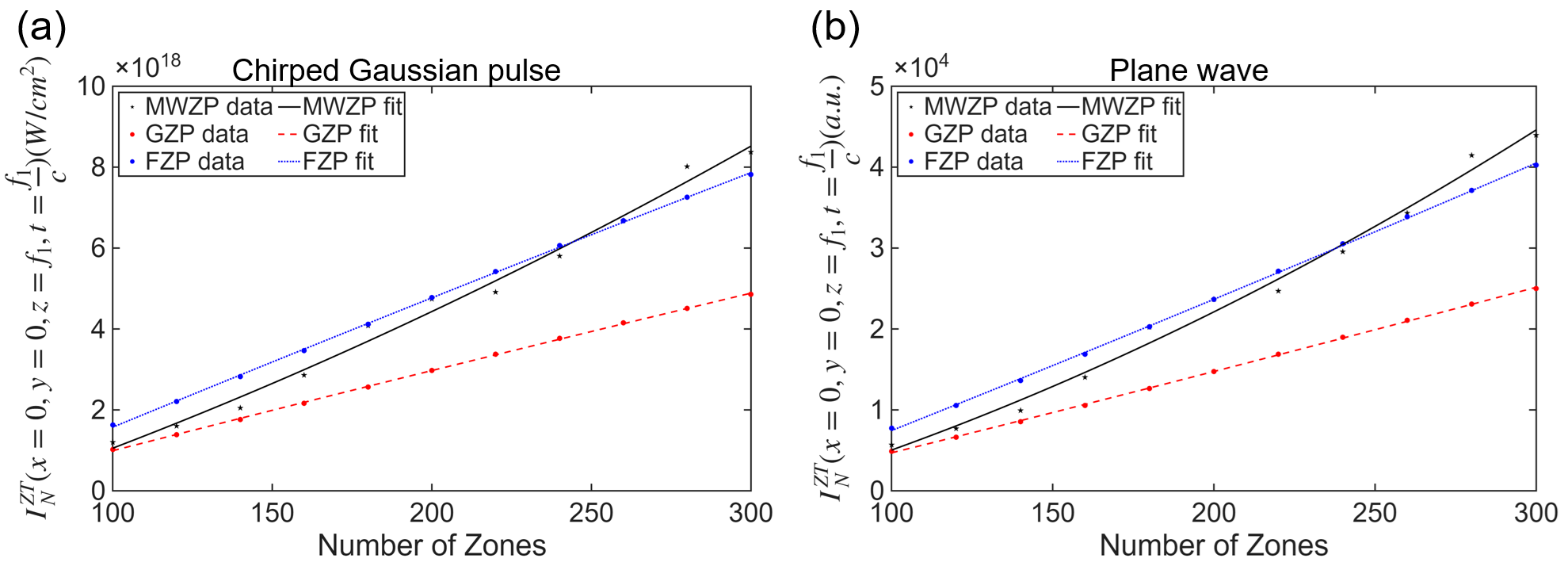}
\caption{The primary focus intensity of MWZPs (black stars), GZPs (red dots), and FZPs (blue dots) as a function of the number of zones. Solid curves represent fits to the corresponding data. (a) The chirped Gaussian pulse case. (b) The plane wave case.
}
\label{changing_plate_number2}
\end{figure}

The dependence of the  intensity $\mathrm{I}_{\mathrm{N}}^{\mathrm{ZT}}$ 
on the number of zone plates ($\mathrm{N}$) is shown in Figure \ref{changing_plate_number2}a and b.
The GZP yields the lowest $\mathrm{I}_{\mathrm{N}}^{\mathrm{ZT}}$ across the studied range $100<\mathrm{N}<300$. 
In contrast, the FZP maintains a strong primary focus. 
The MWZP demonstrates a parabolic increase in its primary focus intensity. 
The fluctuations observed at specific zone numbers are attributed to the sinusoidal width modulation profile. 
 For comparison, the intensity $\mathrm{I}_{\mathrm{N}}^{\mathrm{ZT}}$ when the incoming beam is a plane wave is illustrated
in Figure \ref{changing_plate_number2}b. 
These results are similar to those of the chirped Gaussian pulse.

\begin{figure}[htbp]
\centering \includegraphics[width=1\linewidth]{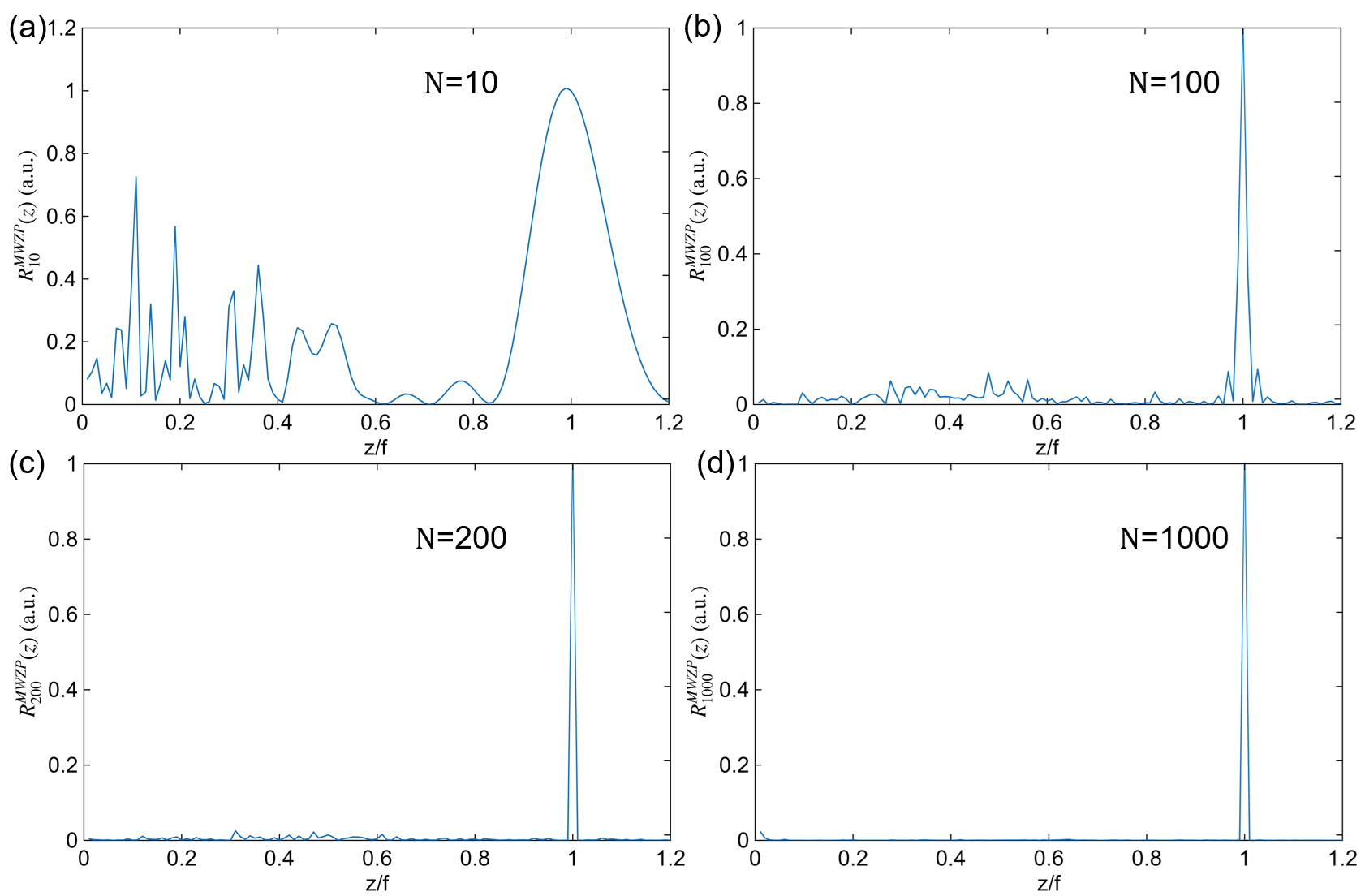}
\caption{Normalized diffraction intensity distribution $\mathrm{R}_{\mathrm{N}}^{\mathrm{MWZP}}(\mathrm{z})$ along the optical axis for MWZPs with different number of zones\cite{ZHANG2018387}. (a) 10 zones. (b) 100 zones. (c) 200 zones. (d) 1000 zones. 
}
\label{multiple_500zone_pairs}
\end{figure}

Furthermore, 
numerical simulations are performed to clarify the dependence of 
the  relative intensity $\mathrm{R}_{\mathrm{N}}^{\mathrm{MWZP}}(\mathrm{z})$ (Eq.\ref{eq.Rn.zt}) and  the full width at half maximum (FWHM) of the primary focal peak  on the number of zone pairs $\mathrm{N}$. 

As shown in Figure \ref{multiple_500zone_pairs}, 
the relative intensities $\mathrm{R}_{\mathrm{N}}^{\mathrm{MWZP}}(\mathrm{z})$ for MWZP with different zone numbers  ($\mathrm{N} = 10, 100, 200$, and 1000) are presented. 
When the MWZP has only N = 10 zones (Figure \ref{multiple_500zone_pairs}a), 
the maximum relative intensity $\mathrm{R}_{\mathrm{N}}^{\mathrm{MWZP}}(\mathrm{z})$ has the order of 0.75 in the non-primary focal range.
The intensity fluctuations  gradually decrease with increasing $\mathrm{N}$. 
Meanwhile, 
the FWHM of the primary focal peak is reduced significantly,  from $\mathrm{FWHM}_{\mathrm{N} = 10} = 8.67$ cm to $\mathrm{FWHM}_{\mathrm{N} = 100} = 0.79$ cm, with increasing  N. 
 For case of N = $200$  (Figure \ref{multiple_500zone_pairs}c), 
 in the non-primary focal range, the maximum relative intensity $\mathrm{R}_{\mathrm{N}}^{\mathrm{MWZP}}(\mathrm{z})$ is only approximately 2\%, and $\mathrm{FWHM}_{\mathrm{N} = 200} = 0.43$ cm; 
and for case of N = 1000 (Figure \ref{multiple_500zone_pairs}d),
they are $<0.6\%$ 
and $\mathrm{FWHM}_{\mathrm{N} = 1000} = 0.10$ cm,  correspondingly.

\subsection{Dependence of chirp type}

As shown in Figure \ref{chirp velocity}, tuning the chirp of incoming laser pulses provides direct control over the flying focus velocity, enabling it to propagate or counter-propagate at arbitrary velocities across the extended focal range. 
By using an MWZP with a radius of 2.8 cm, and 960 zone pairs,
this effect is simulated.

\subsubsection{Flying focus with a positive chirp}

\begin{figure}[htbp]
\centering \includegraphics[width=1\linewidth]{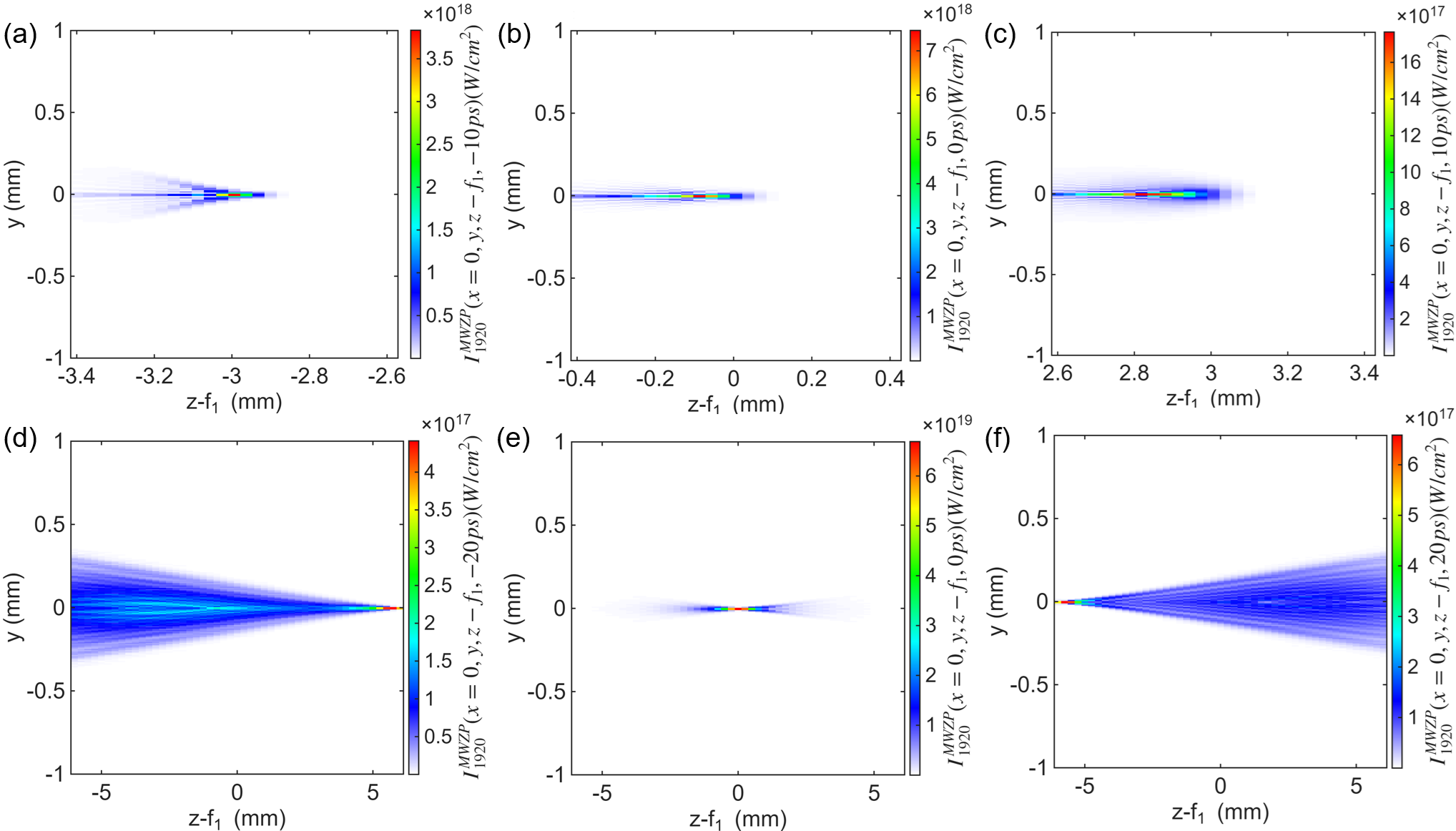}
\caption{Simulated evolution of the flying focus intensity $\mathrm{I}_{1920}^{\mathrm{MWZP}}(\mathrm{x}=0,\mathrm{y},\mathrm{z}-\mathrm{f}_1,\mathrm{t}-\frac{\mathrm{f}_1}{\mathrm{c}})$ for a 115 fs chirped Gaussian pulse with (a-c) positive and (d-f) negative chirp. Time zero corresponds to the arrival of the 800 nm central wavelength at the focus in the negative chirped case. The positively chirped pulse generates a focus that moves forward at a velocity of 0.97c, traversing the focus range from z $\thicksim$ -3 mm to z $\thicksim$ 2.83 mm between -10 ps and 10 ps. Conversely, the negatively chirped pulse produces a focus that propagates backward at a velocity of -1.0c, sweeping from z $\thicksim$ 6 mm to z $\thicksim$ -6 mm between -20 ps and 20 ps.
}
\label{positive and negative chirp}
\end{figure}

Figure \ref{positive and negative chirp}a-c shows the simulation results of the flying focus dynamics for a positively chirped Gaussian pulse. 
In Figure \ref{positive and negative chirp}a,  
 where  $\mathrm{t} = -10$ ps ,
the chirped pulse is  focused at $\mathrm{z} = -3$ mm;
 $\mathrm{t} = 0$ ps, focused at $\mathrm{z} = -0.08$ mm (Figure \ref{positive and negative chirp}b);
and then $\mathrm{t} = 10$ ps, focused at $\mathrm{z} = 2.83$ mm (Figure \ref{positive and negative chirp}c).
The speed of the focal spot is then calculated to be  0.97c,  which is the same as the velocity obtained in Figure \ref{chirp velocity}.
During the process, the focal spot size in the transverse direction, $\sigma_\perp$, maintains constant.

\subsubsection{Flying focus with a negative chirp}

 Figure \ref{positive and negative chirp}d-f illustrates the flying focus dynamics for a negatively chirped Gaussian pulse. 
 The first frame (\ref{positive and negative chirp}d)  shows that the laser reaches a focus at  $\mathrm{z}\simeq6$ mm, which means that the shortest-wavelength component of the chirped Gaussian pulse focuses first. 
Then, the focus moves backward to   $\mathrm{z}\simeq0$ mm (Figure \ref{positive and negative chirp}e), which is a longer-wavelength component of the beam.
Subsequently, the longest wavelength is focused at $\mathrm{z}\simeq -6$ mm, as illustrated in Figure \ref{positive and negative chirp}f. 
The total focal length is about  $\Delta \mathrm{z}\simeq12$ mm, which takes time of about  $\Delta \mathrm{t}\simeq40$ ps, corresponding to a negative velocity of -1.0c. 
The same as the positive chirp case, the focal spot size in the transverse direction of the negative chirp case, $\sigma_\perp$, maintains constant in the propagation process.

\section{Discussion}
\begin{figure}[htbp]
\centering \includegraphics[width=1\linewidth]{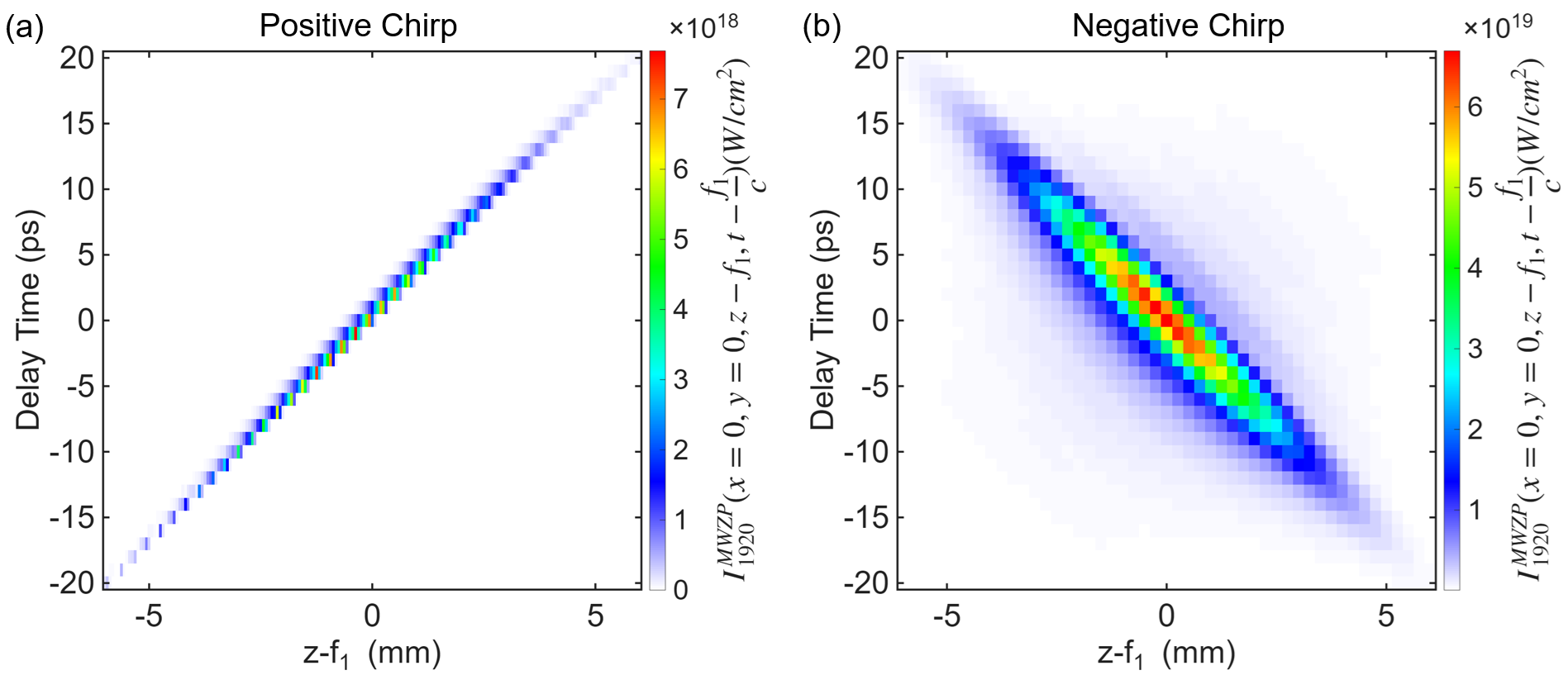}
\caption{
The beam intensity profiles, $\mathrm{I}_{1920}^{\mathrm{MWZP}}(\mathrm{x}=0,\mathrm{y}=0,\mathrm{z}-\mathrm{f}_1,\mathrm{t}-\frac{\mathrm{f}_1}{\mathrm{c}})$,   of the optimized MWZP, showing their dependence on both axial position  $z$ and time $t$. 
The   Gaussian pulse which has width of 115 fs, wavelength of 800 nm,  and   positive  (a)  or   negative chirp (b) are shown.
The  flying focus speed of $+0.97\mathrm{c}$ (a) and $-1.0\mathrm{c}$ (b) can be obtained.
}
\label{Discussion}
\end{figure}

The flying focus scheme has two  important characteristics: the decoupling of focal velocity from the group velocity and extended focal range, 
which can be recognised in Figure \ref{Discussion}, where the 2D plot of  the intensity $\mathrm{I}_{1920}^{\mathrm{MWZP}}(\mathrm{x}=0,\mathrm{y}=0,\mathrm{z}-\mathrm{f}_1,\mathrm{t}-\frac{\mathrm{f}_1}{\mathrm{c}}) $ is shown. 

In Figure \ref{Discussion}(a), 
with a positive chirp, the focus moves forward at a velocity of $+0.97\mathrm{c}$, 
sweeping the range of   $-6<\mathrm{z}< 6$ mm. 
Conversely, in Figure \ref{Discussion}(b), with a negative chirp, the focus flies backward at -1.0c, covering the same range, $-6<\mathrm{z}<6$ mm.
Both of them exhibit the decoupling of the focal velocity from the group velocity. 
By appropriately arranging the chirp of the incident light pulses, the focal spot can propagate at any velocity.
The extended focal region is about 12 mm, exceeding 140 Rayleigh ranges.

As a binary optical element, the MWZP offers another advantage: it is highly compatible with modern nanofabrication, which requires only a single lithography step followed by metal deposition and lift-off, thereby avoiding complex processes, such as etching and multilayer alignments. Therefore, in the experiment, when the required diffractive lense's radius is too large to be obtained in the market or the lense is destroyed by a strong laser, the binary MWZP is an ideal alternative as more zone pairs can be easily added during the fabrication process.

Future work will be devoted to combining the idea of the MWZP with the phase plate, which achieves higher peak intensities with the same number of zone plates. As the focusing efficiency is defined as the integral intensity of the focal spot to the total intensity incident on the entire zone plate, MWZPs, which block part of the incident light as binary optical elements, suffer from a low focusing efficiency of approximately 6.25$\%$\cite{ZHANG2018387}. Meanwhile, although the n-level phase plate is unable to suppress ($\mathrm{n}+1$)th order focus\cite{DiFabrizio1999}, the primary focus of the phase plate has a higher intensity. If the same procedure of simulating the transmittance of GZPs is applied to the phase plate, a phase-modulated MWZP with a higher efficiency can be realized, which also suppresses high-order foci.

\section{Summary}

 In this study,  by modulating the zone width, the number of zone plates, and the chirp parameters of incident Gaussian pulses, 
a modulated width zone plate (MWZP) has been designed for  flying focus applications, which is characterized by deeply suppressed high-order foci,  enhanced peak intensities, and a controllable focal speed.

Numerical simulations confirm that 
the intensities of high-order foci up to the 25th order are suppressed by more than two orders of magnitude with an MWZP of 200 zones.
This single-focus characteristic becomes pronounced when sufficient annular zones are illuminated, which suppresses high noises far from the primary focus. 
With sufficient zone pairs, the main focus intensity is also amplified significantly when compared to the Gabor zone plates (GZPs)  and Fresnel zone plates (FZPs).
With an appropriate chirp parameter for the incident Gaussian pulse, the flying focus can propagate with any forward or backward velocity, whatever smaller than speed of light or larger than the speed of light,  with an extended focal region over 100 Rayleigh ranges.

 The proposed MWZP method may offer significant potential across multiple areas, including 
 high-efficiency laser amplifiers\cite{PhysRevLett.120.024801}, 
 TeV laser wakefield accelerators\cite{10.1063/5.0274780,PhysRevLett.124.134802}, 
 monochromatic EUV sources\cite{ComptonZhao2025},
 and more.

\begin{backmatter}
\bmsection{Funding Statement}
National Natural Science Foundation of China (12235003); the National Key R$\&$D Program of China (2023YFA1606900, 2022YFA1602301).

\bmsection{Data Access Statement}
Data underlying the results presented in this paper are not publicly available at this time but may be obtained from the authors upon reasonable request.

\bmsection{Conflict of Interest declaration}
The authors declare that they have no conflicts of interest.
\end{backmatter}






\end{CJK*}
\end{document}